\documentclass{PoS}
\def\ltap{\raisebox{-.6ex}{\rlap{$\,\sim\,$}} \raisebox{.4ex}{$\,<\,$}} 
\def\ptmin{p_{T{\rm min}}}
\def\ptmax{p_{T{\rm max}}}
\def\ptveto{p_T^{\rm veto}}

\newcommand\as{\alpha_{\mathrm{S}}} 

\title{HNNLO: a Monte Carlo program to compute Higgs boson production at hadron colliders}

\ShortTitle{HNNLO: a MC program to compute Higgs boson production at hadron colliders}

\author{Stefano Catani,\,\,  \speaker{Massimiliano Grazzini}\\
        INFN, Sezione di Firenze, Via Sansone 1, I-50019 Sesto Fiorentino, Florence, Italy\\
}

\abstract{We consider Higgs boson production through gluon--gluon fusion in hadron collisions. We present a numerical program that computes the cross section up to NNLO in QCD perturbation theory. The program includes the decay modes $H\to\gamma\gamma$, $H\to WW\to l\nu l\nu$,
$H\to ZZ\to 4$ leptons,
and allows the user to apply arbitrary cuts on the momenta of
the partons and of the photons or leptons that are
produced in the final state.}

\FullConference{8th International Symposium on Radiative Corrections (RADCOR)\\
		 October 1-5 2007\\
		 Florence, Italy}

\begin{document}

\section{INTRODUCTION}

Gluon-gluon fusion is the main production channel of
the Standard Model Higgs boson at the LHC.
At leading order (LO) in QCD perturbation theory, the cross section is proportional
to $\as^2$, $\as$ being the QCD coupling. The QCD radiative corrections to the total cross section are known at the
next-to-leading order (NLO) 
\cite{Dawson:1990zj,Djouadi:1991tka,Spira:1995rr}
and at the next-to-next-to-leading order (NNLO) 
\cite{Harlander:2000mg,Catani:2001ic,Harlander:2001is,Harlander:2002wh,Anastasiou:2002yz,Ravindran:2003um}.
The effects of a jet veto on the total cross section has been studied
up to NNLO \cite{Catani:2001cr}.
We recall that all the results at NNLO have been obtained by using 
the large-$M_t$ approximation, $M_t$ being the mass of the top quark.

These NNLO calculations can be supplemented with soft-gluon resummed calculations at next-to-next-to-leading logarithmic (NNLL) accuracy
either to improve the quantitative accuracy of the perturbative predictions
(as in the case of the total cross section \cite{Catani:2003zt,Moch:2005ky}) or to provide
reliable predictions in phase-space regions where fixed-order calculations
are known to fail (as in the case of the $q_T$ distribution
of the Higgs boson \cite{Bozzi:2003jy,Bozzi:2005wk,Bozzi:2007pn}
at small $q_T$).

A common feature of these NNLO and NNLL calculations is that they
are fully inclusive over the produced final state (in particular, over
final-state QCD radiation).
Therefore they refer to situations
where the experimental cuts
are either ignored (as in the case of the total cross section) or taken into account
only in simplified cases (as in the case of the jet vetoed cross section).
The impact of higher-order corrections
may be strongly dependent on the details of the applied cuts and also the shape of various
distributions is typically affected by these details.

The first NNLO calculation that fully takes into account experimental cuts
was reported in Ref.~\cite{Anastasiou:2005qj}, considering the decay mode $H\to\gamma\gamma$ (see also \cite{Stockli:2005hz}).
In Ref.~\cite{Anastasiou:2007mz} the calculation is
extended to the decay mode $H\to WW\to l\nu l\nu$ (see also \cite{Anastasiou:2008ik}).

In Ref.~\cite{Catani:2007vq} we have
proposed a method to perform NNLO calculations and we have
applied it to perform an independent computation
of the Higgs production cross section.
The method is completely different from
that used in Refs.~\cite{Anastasiou:2005qj,Anastasiou:2007mz}.
Our calculation is implemented
in a fully-exclusive parton level event generator.
This feature makes it particularly suitable for practical applications
to the computation of distributions in the form of bin histograms.
Our numerical program can be downloaded from \cite{hnnlo}.
The decay modes that are currently implemented
are $H\to \gamma\gamma$ \cite{Catani:2007vq}, $H\to WW\to l\nu l\nu$ and $H\to ZZ\to 4$ leptons \cite{Grazzini:2008tf}.

In the following
we present a brief selection of results that can be obtained
by our program.
We consider Higgs boson production at the LHC and
use the MRST2004 parton distributions \cite{Martin:2004ir},
with parton densities and $\as$ evaluated at each corresponding order
(i.e., we use $(n+1)$-loop $\as$ at N$^n$LO, with $n=0,1,2$). The 
renormalization and factorization scales are fixed to the value 
$\mu_R=\mu_F=M_H$, where $M_H$ is the mass of the Higgs boson.

\section{RESULTS FOR THE DECAY MODE $H\to \gamma\gamma$}

\begin{figure}
\begin{center}
\includegraphics[width=0.5\textwidth]{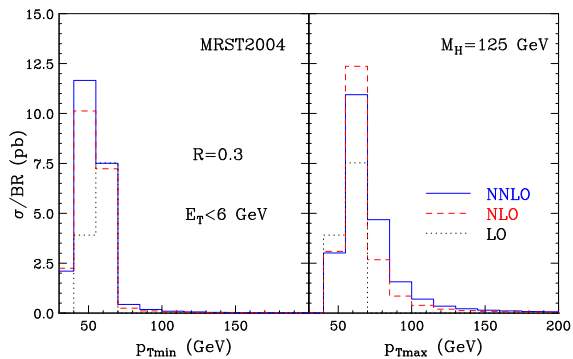}
 \caption{Distributions in $\ptmin$ and $\ptmax$ for the diphoton signal at the LHC. The cross section is divided by the branching ratio in two photons.}
\label{fig:isol}
\end{center}
\end{figure}

We consider the production of a Higgs boson of mass $M_H=125$ GeV in the
$H\to\gamma\gamma$ decay mode and follow
Ref.~\cite{CMStdr} to apply cuts on the photons.
For each event, we classify the photon transverse momenta according to their
minimum and maximum value,  
$\ptmin$ and $\ptmax$. The photons are required to be in the
central rapidity region, $|\eta|<2.5$, with  $\ptmin>35$~GeV
and $\ptmax>40$~GeV. We also require the photons to be isolated:
the hadronic (partonic) transverse energy in a cone of radius $R=0.3$ along the
photon direction 
has to be smaller than 6~GeV. Using these cuts
the impact of the NNLO corrections on the NLO total cross section
is reduced from 19\% to 11\%.
 
In Fig.~\ref{fig:isol} we plot
the distributions in $\ptmin$ and $\ptmax$
of the signal process $gg\to H\to\gamma\gamma$.
We note that the shape of these distributions sizeably
differs when going from LO to NLO and to NNLO.
The origin of these perturbative instabilities is well known 
\cite{Catani:1997xc}.
Since the LO spectra
are kinematically bounded by $p_T\leq M_H/2$,
each higher-order perturbative contribution produces
(integrable) logarithmic singularities in the vicinity of
that boundary. More detailed studies are necessary to assess
the theoretical uncertainties of these fixed-order results
and the relevance of all-order resummed calculations.
 
\begin{figure}
\begin{center}
\includegraphics[width=0.5\textwidth]{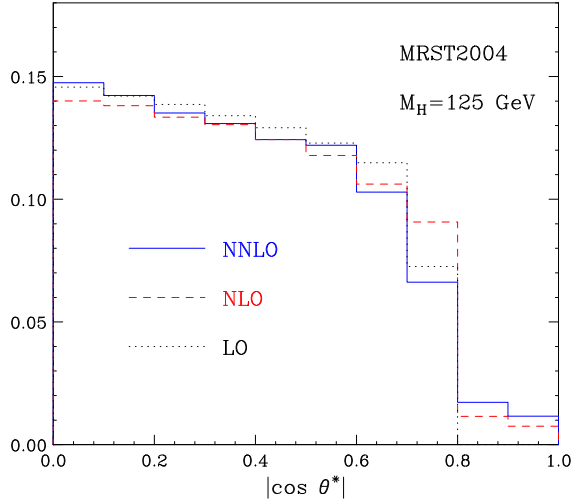}
 \caption{Normalized distribution in the variable $\cos\theta^*$.}
\label{fig:ctheta}
\end{center}
\end{figure}

In Fig.~\ref{fig:ctheta} we consider the (normalized) distribution in
the variable $\cos\theta^*$, where $\theta^*$ is the polar angle of one of the photons
in the rest frame of the Higgs boson
\footnote{We thank Suzanne Gascon and Markus Schumacher for suggesting
the use of this variable.}.
At small values of $\cos\theta^*$ the distribution
is quite stable with respect to higher order QCD corrections.
We also note that the LO distribution vanishes beyond the value $\cos\theta^*_{\rm max}<1$.
The upper bound $\cos\theta^*_{\rm max}$ is due to the fact that the photons
are required to have a minimum $p_T$ of $35$ GeV.
As in the case of Fig.~\ref{fig:isol}, in the vicinity of this LO kinematical boundary
there is an instability of the
perturbative results beyond LO.

\section{RESULTS FOR THE DECAY MODE $H\to l\nu l\nu$}

We now consider the production of a Higgs boson with mass $M_H=165$ GeV in the decay mode $H\to l\nu l\nu$.
We apply a set of {\em preselection} cuts taken from the study of Ref.~\cite{Davatz:2004zg}. The charged leptons have $p_T$ larger than 20 GeV, and $|\eta|<2$. The missing $p_T$ is
larger than $20$ GeV and the invariant mass of the charged
leptons is smaller than $80$ GeV.
Finally, the azimuthal separation of the charged leptons in the
transverse plane ($\Delta\phi$) is smaller than $135^o$.
By applying these cuts the impact of the NNLO corrections
on the NLO result does not change and is of about $20\%$.

\begin{figure}
\begin{center}
\includegraphics[width=0.5\textwidth]{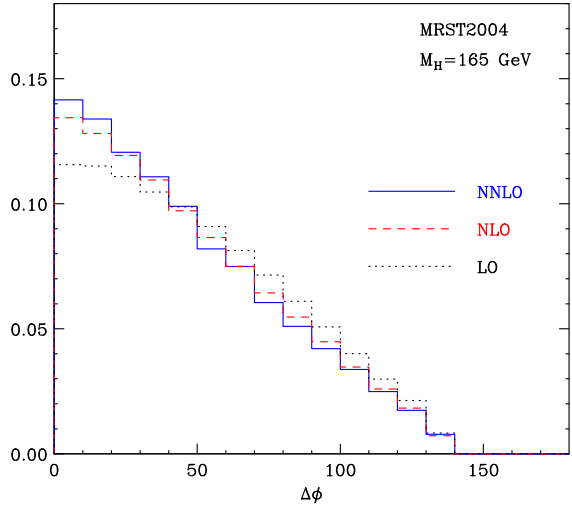}
 \caption{Normalized $\Delta\phi$ distribution at LO, NLO, NNLO.}
\label{fig:deltaphi}
\end{center}
\end{figure}
In Fig.\ref{fig:deltaphi} we
plot the $\Delta\phi$ distribution at LO, NLO and NNLO.
As is well known \cite{Dittmar:1996ss},
the charged leptons from the Higgs boson
signal tend to be close in angle,
and thus the distribution is peaked at small $\Delta\phi$.
We note that the effect of the QCD corrections is to increase
the steepness of the distribution,
from LO to NLO and from NLO to NNLO.

We finally apply the set of selection cuts designed to isolate the Higgs boson signal \cite{Davatz:2004zg}.
The transverse momenta of the two charged leptons
are classified according to their
minimum and maximum value,  
$\ptmin$ and $\ptmax$.
They should fulfil $\ptmin > 25$ GeV and $35~{\rm GeV}< \ptmax <50$ GeV.
The missing $p_T$ of the event should be larger than 20 GeV
and the invariant mass of the charged leptons should be smaller than $35$ GeV.
To exploit the steepness of the $\Delta\phi$ distribution shown in Fig.~\ref{fig:deltaphi},
$\Delta\phi$ should be smaller than $45^o$.
Finally, there should be no jets with $p_T^{\rm jet}$ larger
than a given value $\ptveto$.

The corresponding cross sections, for different values of $\ptveto$, are reported in Table ~\ref{table}.

\begin{table}[htbp]
\begin{center}
\begin{tabular}{|c|c|c|}
\hline
$\ptveto$ (GeV) & $\sigma_{NLO}$ (fb) & $\sigma_{NNLO}$ (fb)\\
\hline
\hline
no veto & $21.26\pm 0.05$ & $22.21\pm 0.32$\\
\hline
40 & $18.62\pm 0.05$ & $17.38\pm 0.34$\\
\hline
30 & $17.18\pm 0.05$ & $15.74\pm 0.35$\\
\hline
20 & $14.42\pm 0.05$ & $11.31\pm 0.38$\\
\hline

\end{tabular}
\end{center}
\caption{{\em Cross sections for $pp\to H+X\to WW+X\to l\nu l\nu+X$ at the LHC
when selection cuts are applied for different $\ptveto$.}}
\label{table}
\end{table}
We see that, even without the jet veto, the impact of radiative corrections is strongly reduced with this choice
of cuts. The jet veto further reduces the effect of the NNLO corrections, which become negative for $\ptveto \ltap 40$ GeV.

\section{SUMMARY}

We have illustrated the results of a calculation of the Higgs boson production cross section at the LHC
up to NNLO in QCD perturbation theory. The calculation is performed
by using the numerical program {\tt HNNLO},
which includes all the relevant decay modes of the Higgs boson, namely,
$H\to\gamma\gamma$, $H\to WW\to l\nu l\nu$ and $H\to ZZ\to 4$ leptons. The program allows the user to apply arbitrary cuts on the momenta of the partons and
of the photons/leptons produced in the final state, and to obtain the required distributions in the form of
bin histograms. We have presented a brief selection of numerical results for the decay modes $H\to\gamma\gamma$ and
$H\to WW\to l\nu l\nu$.  More detailed results for the decay modes $H\to WW$ and $H\to ZZ$
can be found in Ref.~\cite{Grazzini:2008tf}.
The fortran code {\tt HNNLO} can be downloaded from \cite{hnnlo}.

\bibliography{radcor}

\begin{thebibliography}{10}

\bibitem{Dawson:1990zj}
S. Dawson,
\newblock Nucl. Phys. B359 (1991) 283.

\bibitem{Djouadi:1991tka}
A. Djouadi, M. Spira and P.M. Zerwas,
\newblock Phys. Lett. B264 (1991) 440.

\bibitem{Spira:1995rr}
M. Spira et~al.,
\newblock Nucl. Phys. B453 (1995) 17, hep-ph/9504378.

\bibitem{Harlander:2000mg}
R.V. Harlander,
\newblock Phys. Lett. B492 (2000) 74, hep-ph/0007289.

\bibitem{Catani:2001ic}
S. Catani, D. de~Florian and M. Grazzini,
\newblock JHEP 05 (2001) 025, hep-ph/0102227.

\bibitem{Harlander:2001is}
R.V. Harlander and W.B. Kilgore,
\newblock Phys. Rev. D64 (2001) 013015, hep-ph/0102241.

\bibitem{Harlander:2002wh}
R.V. Harlander and W.B. Kilgore,
\newblock Phys. Rev. Lett. 88 (2002) 201801, hep-ph/0201206.

\bibitem{Anastasiou:2002yz}
C. Anastasiou and K. Melnikov,
\newblock Nucl. Phys. B646 (2002) 220, hep-ph/0207004.

\bibitem{Ravindran:2003um}
V. Ravindran, J. Smith and W.L. van Neerven,
\newblock Nucl. Phys. B665 (2003) 325, hep-ph/0302135.

\bibitem{Catani:2001cr}
S. Catani, D. de~Florian and M. Grazzini,
\newblock JHEP 01 (2002) 015, hep-ph/0111164.

\bibitem{Catani:2003zt}
S. Catani et~al.,
\newblock JHEP 07 (2003) 028, hep-ph/0306211.

\bibitem{Moch:2005ky}
S. Moch and A. Vogt,
\newblock Phys. Lett. B631 (2005) 48, hep-ph/0508265.

\bibitem{Bozzi:2003jy}
G. Bozzi et~al.,
\newblock Phys. Lett. B564 (2003) 65, hep-ph/0302104.

\bibitem{Bozzi:2005wk}
G. Bozzi et~al.,
\newblock Nucl. Phys. B737 (2006) 73, hep-ph/0508068.

\bibitem{Bozzi:2007pn}
G. Bozzi et~al.,
\newblock Nucl. Phys. B791 (2008) 1, arXiv:0705.3887 [hep-ph].

\bibitem{Anastasiou:2005qj}
C. Anastasiou, K. Melnikov and F. Petriello,
\newblock Nucl. Phys. B724 (2005) 197, hep-ph/0501130.

\bibitem{Stockli:2005hz}
F. Stockli, A.G. Holzner and G. Dissertori,
\newblock JHEP 10 (2005) 079, hep-ph/0509130.

\bibitem{Anastasiou:2007mz}
C. Anastasiou, G. Dissertori and F. Stockli,
\newblock JHEP 09 (2007) 018, arXiv:0707.2373 [hep-ph].

\bibitem{Anastasiou:2008ik}
C. Anastasiou et~al.,
\newblock arXiv:0801.2682 [hep-ph].

\bibitem{Catani:2007vq}
S. Catani and M. Grazzini,
\newblock Phys. Rev. Lett. 98 (2007) 222002, hep-ph/0703012.

\bibitem{hnnlo}
{\tt http://theory.fi.infn.it/grazzini/codes.html} .

\bibitem{Grazzini:2008tf}
M. Grazzini,
\newblock arXiv:0801.3232 [hep-ph].

\bibitem{Martin:2004ir}
A.D. Martin et~al.,
\newblock Phys. Lett. B604 (2004) 61, hep-ph/0410230.

\bibitem{CMStdr}
CMS Physics, Technical Design Report, Vol.~II Physics Performance , report
  CERN/LHCC 2006-021.

\bibitem{Catani:1997xc}
S. Catani and B.R. Webber,
\newblock JHEP 10 (1997) 005, hep-ph/9710333.

\bibitem{Davatz:2004zg}
G. Davatz et~al.,
\newblock JHEP 05 (2004) 009, hep-ph/0402218.

\bibitem{Dittmar:1996ss}
M. Dittmar and H.K. Dreiner,
\newblock Phys. Rev. D55 (1997) 167, hep-ph/9608317.

\end{thebibliography}

\end{document}